\documentstyle[multicol,aps,prb,epsf]{revtex} 

\begin{document}
\title{Anomalous crossover between thermal and shot noise in macroscopic diffusive
conductors}
\author{G. Gomila and L. Reggiani}
\address{Dipartimento di Ingegneria dell' Innovazione and \\
Istituto Nazionale di Fisica della Materia, Universit\'a di Lecce, Via\\
Arnesano s/n, 73100-Lecce, Italy}
\maketitle

\begin{abstract}
We predict the existence of an anomalous crossover between thermal and shot
noise in macroscopic diffusive conductors. We first show that, besides
thermal noise, these systems may also exhibit shot noise due to fluctuations
of the total number of carriers in the system. Then we show that at
increasing currents the crossover between the two noise behaviors is
anomalous, in the sense that the low frequency current spectral density
displays a region with a superlinear dependence on the current up to a cubic
law. The anomaly is due to the non-trivial coupling in the presence of the
long range Coulomb interaction among the three time scales relevant to the
phenomenon, namely, diffusion, transit and dielectric relaxation time.
\end{abstract}

\begin{multicols}{2}
\narrowtext
Shot noise and thermal noise are the two prototypes of noise present in
nature.\cite{Ziel85,Kogan96} Thermal noise is displayed by a conductor at or
near equilibrium, and is associated with its conductance through Nyquist
theorem\cite{Nyquist28} $S_{I}^{ther}(0)=4k_{B}TG$, where $S_{I}^{ther}(0)$
is the low frequency current spectral density, $k_{B}$ the Boltzman
constant, $T$ the temperature and $G$ the conductance. Shot noise is due to
the discreteness of the carriers charge, and displays a low frequency
spectral density of current fluctuations in the form $S_{I}^{shot}(0)=\gamma
2q\overline{I}$, where $\overline{I}$ is the average dc current, $q$ the
carrier charge and $\gamma $ the so called Fano Factor. Being an excess
noise, it can only be observed under non-equilibrium conditions and provides
information not available from linear response coefficients such as
conductance. Following Landauer's ideas, \cite{Landauer92} these two types
of noise are special forms of a more general noise formula representing
different manifestations of the same underlying microscopic mechanisms. As a
result, for systems displaying shot noise one should expect a continuous and
smooth transition between the equilibrium thermal noise and the
non-equilibrium shot noise. Two examples of such transitions are provided by
the expressions 
\begin{equation}
S_{I}(0)=2q\bar{I}\coth \left( \frac{qV}{2k_{B}T}\right) \text{,}
\label{PIintro}
\end{equation}
and 
\begin{equation}
S_{I}(0)=4k_{B}TG\left[ (1-\gamma )+\gamma \frac{qV}{2k_{B}T}\coth \left( 
\frac{qV}{2k_{B}T}\right) \right] \text{,}  \label{PintroQ}
\end{equation}
which represent standard transitions for a classical and a quantum system,
respectively.\cite{Ziel85,Landauer92} In previous equations $V$ is the
applied voltage. In both cases one obtains $S_{I}^{ther}(0)$ at or near
equilibrium, when $\left| qV/k_{B}T\right| \ll 1$, and $S_{I}^{shot}(0)$ far
from equilibrium, when $\left| qV/k_{B}T\right| \gg 1$.

A variety of classical and quantum physical systems exhibit the above $coth$%
-like cross-over. Among them we remind $p-n$ junctions\cite{Ziel85},
Schottky barrier diodes\cite{Ziel85}, tunnel diodes\cite{Ziel85}, mesoscopic
diffusive conductors with coherent\cite{deJong96}, and semiclassical
transport \cite{Nagaev92,Gonzalez99}, etc. We remark that an essential
feature of the above formulae is to predict a monotonic increase of the
spectral density with current which never exceeds a linear dependence.
Finally, we note that it is common believe that macroscopic conductors do
not display shot noise.\cite{Shimizu92}

The aim of this article is to prove that macroscopic conductors can display
shot noise and that the transition between thermal and shot noise shows a
remarkable deviation from the standard $coth$-like behavior. In particular,
the region of cross-over evidences a current spectral density which
increases more than linearly with current, up to a cubic dependence.

The system under consideration is a macroscopic homogeneous diffusive
conductor of length $L$, (henceforth shortly referred to as {\em macroscopic
diffusive conductor}). The conductor is considered to be macroscopic in the
sense that the sample length $L$ satisfies $L\gg l_{in},l_{e}$, where $%
l_{in} $ and $l_{e}$ are the inelastic and elastic mean free paths,
respectively. Moreover, homogeneous conditions implies that the stationary
electric field and charge density profiles are homogeneous. Although at
first sight it seems surprising that macroscopic diffusive conductors are
able to display shot noise, see for instance Ref. \onlinecite{Shimizu92}, it
is easy to convince oneself that this is indeed the case. The key argument
is provided by the fact that the diffusion of carriers through the sample, a
part from velocity fluctuations, also induce {\em fluctuations of the total
number of particles inside the sample}. These number fluctuations are
related to the fact that the time a carrier spends to cross the sample
depends on the particular succession of scattering events, thus giving rise
to fluctuations in the instantaneous value of the total number of particles
inside the sample. As a consequence, besides the usual thermal noise
associated with velocity fluctuations, we will have an excess noise
associated with number fluctuations. Note, that existing arguments against
the presence of shot noise in macroscopic conductors are always based on the
assumption that number fluctuations are negligible, what is not always true
in macroscopic diffusive conductors, as will be shown below.

That number fluctuations can give rise to shot noise can be seen as follows.
The excess noise associated with number fluctuations can be characterized as%
\cite{Kuhn90} 
\begin{equation}
S_{I}^{ex}(0)=\left( \frac{\overline{I}}{\overline{N}}\right) ^{2}S_{N}(0)%
\text{,}  \label{Pexintro}
\end{equation}
where $S_{N}(0)=2\int_{-\infty }^{+\infty }dt\overline{\delta N(0)\delta N(t)%
}$ is the low frequency spectral density of number fluctuations and $%
\overline{N}$ the average number of carriers inside the system. Furthermore,
within an exponential model for the decay of number fluctuations one assumes 
$S_{N}(0)=\overline{\delta N^{2}}\tau _{N}$ where $\overline{\delta N^{2}}$
is the variance and $\tau _{N}$ the relaxation time for such a fluctuations.%
\cite{remark1} If the relaxation of number fluctuations takes place on a
time scale of the order of the transit time $\tau _{T}$, then one has $\tau
_{N}\sim \tau _{T}$. By using in Eq.(\ref{Pexintro}) that for a diffusive
conductor $\tau _{T}=L/v=q\overline{N}/\overline{I}$, where $v$ is the drift
velocity, and where we have used that $\overline{I}=qA\overline{n}v$, with $%
A $ being the cross sectional area and $\overline{n}$ the average carrier
density, we obtain $S_{I}^{ex}(0)\sim q\left( \overline{\delta N^{2}}/%
\overline{N}\right) \overline{I}$, which is shot noise like.

Therefore, macroscopic diffusive conductors offer a new and simple example
in which to investigate in detail the transition between thermal and shot
noise. To this purpose, we need an explicit expression for the current
spectral density valid, in particular, in the transition region between
thermal and shot noise. This explicit expression can be obtained by solving
the appropriate equations for the fluctuations. For simplicity the sample is
assumed to have a transversal size sufficiently thick to allow a one
dimensional electrostatic treatment in the $x$ direction and to neglect the
effects of boundaries in the $y$ and $z$ directions. Furthermore, since we
are interested in the low frequency noise properties (beyond $1/f$ noise),
we will neglect the displacement current. Accordingly the standard
drift-diffusion Langevin equation for a macroscopic diffusive conductor reads%
\cite{VanVliet94} 
\begin{equation}
\frac{I(t)}{A}=qn\mu E+qD\frac{dn}{dx}+\frac{\delta I_{x}(t)}{A}\text{,}
\label{DriftDif}
\end{equation}
which after linearization around the stationary homogeneous state gives\cite
{remark2} 
\begin{eqnarray}
\frac{\delta I(t)}{A} &=&q\mu \overline{E}\eta \delta n_{x}(t)+q\overline{n}%
\mu \delta E_{x}(t)+  \label{DrifDifFluc} \\
&&qD\frac{d\delta n_{x}(t)}{dx}+\frac{\delta I_{x}(t)}{A}\text{.}  \nonumber
\end{eqnarray}
Here, $\delta E_{x}(t)$ and $\delta n_{x}(t)$ refer to the fluctuations of
electric field and number density at point $x$, respectively, while $\delta
I(t)$ refers to the fluctuations of the total current. Moreover, $\mu $ is
the mobility, $\bar{E}$ the average electric field, $D$ the diffusion
coefficient and the bar denotes a time average. We assume that $\mu $ and $D$
may depend on $\overline{n}$, in order to include in the model also
degenerate conductors. The numerical factor $\eta =\left( 1+\frac{\mu
_{N}^{\prime }}{\mu /\overline{n}}\right) $ , with $\mu _{N}^{\prime }=\frac{%
\partial \mu }{\partial \overline{n}}$, accounts for the possible dependence
of the mobility on the number density and $\delta I_{x}(t)$ is a Langevin
noise source, which accounts for the fluctuations of current due to the
diffusion of carriers inside the sample. It has zero mean and correlation
function, 
\begin{equation}
\left\langle \delta I_{x}(t)\delta I_{x^{^{\prime }}}(t^{\prime
})\right\rangle =\frac{1}{2}K\delta (x-x^{\prime })\delta (t-t^{\prime })%
\text{,}  \label{Correlation}
\end{equation}
where $K=4qAk_{B}T\mu \overline{n}$ is the strength of the fluctuations.
Equation (\ref{DrifDifFluc}) must be supplemented with the Poisson equation 
\begin{equation}
\frac{d\delta E_{x}(t)}{dx}=-\frac{q}{\epsilon }\delta n_{x}(t)\text{,}
\label{Poison}
\end{equation}
where $\epsilon $ is the electric permittivity. Generally, Eqs.(\ref
{DrifDifFluc}) and (\ref{Poison}) are combined into a single equation for
the electric field fluctuation of the form

\begin{equation}
\left( \frac{d^{2}}{dx^{2}}+\frac{1}{L_{E}}\frac{d}{dx}-\frac{1}{L_{D}^{2}}%
\right) \delta E_{x}(t)=\frac{(\delta I_{x}(t)-\delta I(t))}{\epsilon AD}%
\text{,}  \label{dExEq}
\end{equation}
where $L_{E}=D/\eta \mu \overline{E}$ and $L_{D}=\left( D\epsilon /\mu q%
\overline{n}\right) ^{1/2}$.Here, $L_{E}/L$ characterizes the ratio between
a characteristic carrier energy and the energy supplied by the applied
voltage, and $L_{D}$ is the Debye screening length. The ratio $L/L_{D}$
constitutes a relevant indicator of the effects of the long range Coulomb
interaction on the current fluctuations, since for $L/L_{D}\ll 1$, one can
neglect the term proportional to $\delta E_{x}(t)$ in Eq.(\ref{DrifDifFluc}%
), and the equation for the current fluctuations becomes uncoupled from the
Poisson equation. Moreover, since contact effects are negligible we will use
as boundary conditions $\delta n_{0}=\delta n_{L}=0$, which gives, 
\begin{equation}
\left. \frac{d\delta E_{x}(t)}{dx}\right| _{0}=\text{ }\left. \frac{d\delta
E_{x}(t)}{dx}\right| _{L}=0\text{.}  \label{dExbc}
\end{equation}
Equation (\ref{dExEq}), together with Eqs. (\ref{Correlation}) and Eq.(\ref
{dExbc}), constitute a complete set of equations to analyze the noise
properties of macroscopic diffusive conductors. In the present form, they
can be used to describe both degenerate as well as non-degenerate
conductors. The fact that the same underlying scattering mechanisms are
responsible for the noise properties of the system is reflected by the
presence of a unique Langevin source in the model. Being Eq.(\ref{dExEq}) a
second order differential equation with constant coefficients, its solution
can be obtained in a closed analytical form. Hence, from the expression of $%
\delta E_{x}(t)$ one can compute the voltage fluctuation under fixed current
conditions $\delta _{I}V(t)=\int_{0}^{L}dx\delta E_{x}(t)$ (where one uses $%
\delta I(t)=0$), from where the current spectral density can be obtained as $%
S_{I}(0)=G^{2}2\int_{-\infty }^{+\infty }dt\overline{\delta V_{I}(0)\delta
_{I}V(t)}$, with $G=qA\mu \overline{n}/L$. After simple but cumbersome
algebra, the final result can be written in the form 
\begin{equation}
S_{I}(0)=S_{I}^{ther}(0)+S_{I}^{ex}(0)\text{,}  \label{PITot}
\end{equation}
where 
\begin{equation}
S_{I}^{ther}(0)=\frac{K}{L}=4k_{B}TG\text{,}  \label{P0}
\end{equation}
and where 
\begin{eqnarray}
S_{I}^{ex}(0) &=&K\frac{(\lambda _{2}^{2}-\lambda _{1}^{2})}{2L^{2}\lambda
_{1}^{2}\lambda _{2}^{2}}\frac{\left( e^{\lambda _{1}L}-1\right) \left(
e^{\lambda _{2}L}-1\right) }{\left( e^{\lambda _{2}L}-e^{\lambda
_{1}L}\right) ^{2}}\times  \nonumber \\
&&\left[ \lambda _{2}\left( e^{\lambda _{2}L}+1\right) \left( e^{\lambda
_{1}L}-1\right) -\right.  \nonumber \\
&&\left. \lambda _{1}\left( e^{\lambda _{1}L}+1\right) \left( e^{\lambda
_{2}L}-1\right) \right] \text{.}  \label{Pex}
\end{eqnarray}
Here, $\lambda _{1}$and $\lambda _{2}$ are the two eigenvalues of Eq.(\ref
{dExEq}) and are given by 
\begin{equation}
\lambda _{1,2}=-\frac{1}{2L_{E}}\left( 1\pm \sqrt{1+4\frac{L_{E}^{2}}{%
L_{D}^{2}}}\right) \text{.}  \label{l1l2}
\end{equation}
Equations (\ref{PITot})-(\ref{Pex})constitute the general expression for the
low frequency current spectral density of a macroscopic diffusive conductor,
and represent the main result of the present paper. In Eq.(\ref{PITot}) we
distinguish two different contributions. The first one, $S_{I}^{ther}(0)$,
corresponds to thermal noise. The second one, $S_{I}^{ex}(0)$, constitutes
an excess noise and it is directly related to carrier number fluctuations.
This can be proved directly by computing $S_{N}(0)$ from the solution of Eq.(%
\ref{dExEq}) by considering that the number fluctuations are given through $%
\delta N(t)=A\int_{0}^{L}dx\delta n_{x}=A\frac{\epsilon }{q}(\delta
E_{0}(t)-\delta E_{L}(t))$. One then obtains the identity 
\begin{equation}
S_{I}^{ex}(0)=\left( \frac{\overline{I}}{\overline{N}}\right) ^{2}\eta
^{2}S_{N}(0)\text{.}  \label{PexPN}
\end{equation}
Equation (\ref{PexPN}) is of the form of Eq.(\ref{Pexintro}) except for the
presence of $\eta $ which accounts for the possible dependence of the
mobility on carrier density. From Eqs.(\ref{Pex}) and (\ref{PexPN}), it can
be shown that when $L_{D}^{2}/L_{E}\gg L\gg L_{D}$ or $L_{D}\gg L\gg L_{E}$
one has 
\begin{equation}
S_{I}^{ex}(0)=2\gamma q\overline{I}\text{,}  \label{PexShot}
\end{equation}
where $\gamma =\eta k_{B}T\frac{\partial \ln \overline{N}}{\partial E_{F}}$.
This result proves the possibility for macroscopic diffusive conductors to
display shot noise. By defining a characteristic time associated to number
fluctuations through $\tau _{N}=S_{N}(0)/\overline{\delta N^{2}}^{eq}$, with 
$\overline{\delta N^{2}}^{eq}=\overline{N}k_{B}T\frac{\partial \ln \overline{%
N}}{\partial E_{F}}$ being the variance of number fluctuation at
equilibrium, Eq.(\ref{PexShot}) corresponds to a situation in which $\tau
_{N}\approx (2/\eta )\tau _{T}$, thus confirming that when number
fluctuations relax on the time scale given by the transit time they give
rise to shot noise.

Now we are in a position to investigate the properties of the transition
between thermal and shot noise. In Fig.\ref{Fig1} we display the current
spectral density for an ohmic conductor obtained from Eqs.(\ref{PITot})-(\ref
{Pex}), as a function of current for different sample lengths. The current
is normalized to $I_{R}=GV_{R}$ where $V_{R}=\frac{\overline{n}}{\eta q}%
\frac{\partial E_{F}}{\partial \overline{n}}$. In the present units the
curves corresponding to $L/L_{D}<1$ are indistinguishable from the curve
corresponding to $L/L_{D}=1$. In the figure we can easily identify the
thermal and shot noise regimes as the constant and proportional to current
behaviors, respectively. Also depicted for comparison is the current
spectral density of Eq.(\ref{PIintro}) that represents the standard
transition between thermal and shot noise for a classical system (empty
squares). Remarkably, while the transition between thermal and shot noise
follows the standard form for $L<L_{D}$, in the opposite case $L>L_{D}$ it
is anomalous. The anomaly is characterized by a spectral density which at
most increases with the third power of the current tending asymptotically to 
\begin{equation}
\frac{S_{I}(0)}{S_{I}^{ther}(0)}=\left[ 1+\frac{1}{2}\left( \frac{L_{D}}{L}%
\right) ^{4}\left( \frac{\overline{I}}{I_{R}}\right) ^{3}\right] \text{,}
\label{PICubic}
\end{equation}
which holds for $0\leq I\lesssim \left( L/L_{D}\right) ^{2}I_{R}$ as can be
seen in Fig.\ref{Fig1}, where the filled circles represent Eq.(\ref{PICubic}%
). Since this anomalous crossover is absent for $L<L_{D}$ , i.e. when the
long range Coulomb interaction does not affect the current fluctuations, we
conclude that this interaction plays a central role in this unexpected
behavior.

To better understand the role of the long range Coulomb interaction in the
origin of this anomaly, we will analyze how the three characteristic times
in the system combine to yield $\tau _{N}$. For the present case the
following characteristic times can be identified: the diffusion time $\tau
_{D}=L^{2}/D$, the dielectric relaxation time, $\tau _{d}=\epsilon /q%
\overline{n}\mu $ and the, already defined, transit time $\tau _{T}$. 

In Fig. \ref{Fig2} we plot $\tau _{N}$ as obtained from our theory as a
function of current for different sample lengths. Here, we clearly identify
two different behaviors for $\tau _{N}$ depending on whether $L/L_{D}\ll 1$
or $L/L_{D}\gg 1$. For $L/L_{D}\ll 1$ we observe a smooth transition between
the equilibrium value $\tau _{N}\approx 1/3\tau _{D}$ and the far from
equilibrium value $\tau _{N}\approx (2/\eta )\tau _{T}$. This result shows
that when the long range Coulomb interaction is not effective, only $\tau
_{D}$ and $\tau _{T}$ are relevant. As a consequence, near equilibrium we
have $\tau _{D}\ll \tau _{T}$ and number fluctuations are governed by
diffusion, while far from equilibrium we have $\tau _{D}\gg \tau _{T}$ and
they are governed by the transit time, thus giving rise to shot noise. On
the other hand, when $L/L_{D}\gg 1$ the transition between the equilibrium
value $\tau _{N}\approx 4(\tau _{d}/\tau _{D})^{1/2}\tau _{d}$ and the far
from equilibrium value $\tau _{N}\approx (2/\eta )\tau _{T}$ is mediated by
a region in which $\tau _{N}\approx 2\eta \tau _{d}^{2}/\tau _{T}$. The far
from equilibrium behavior, being dominated by the transit time gives rise to
shot noise, while in the intermediate region $\tau _{N}$ is proportional to
the current thus giving rise to the cubic dependence of the current spectral
density. Notice that the transition between the intermediate and the shot
noise region takes place when $\tau _{N}\sim \tau _{d}\sim \tau _{T}$. From
these results we conclude that the origin of the anomalous transition
between thermal and shot noise can be found in the non-trivial coupling
between the different characteristic times in the presence of long range
Coulomb interaction.

From the previous analysis we argue that there are two possible ways of
providing an experimental test of our theory. The first way is an indirect
test to be performed at or neat equilibrium. It consists in proving the
non-trivial coupling of the characteristic times in the presence of the long
range Coulomb interaction. In this case, when $L/L_{D}\gg 1$, one should
obtain a characteristic time for number fluctuations in agreement with the
relationship $\tau _{N}\approx 4\left( \tau _{d}/\tau _{D}\right) ^{1/2}\tau
_{d}=4(\epsilon /q\mu \overline{n})^{3/2}D^{1/2}/L$. The second way is a
direct test, which consists in observing the current dependence of the
current spectral density. According to Eq.(\ref{PICubic}) one should observe
the anomalous transition for $L\gg L_{D}$ when $\overline{I}\gtrsim
(L/L_{D})^{4/3}I_{R}$, or analogously for $\overline{V}/L=\overline{E}
\gtrsim (L/L_{D})^{4/3}V_{R}/L$. To this end, non-degenerate semiconductor
systems offer the best possibilities. For a non-degenerate semiconductor,
with typical parameters $\overline{n}\sim 10^{14}cm^{-3}$, $T\sim 300K$, $%
\epsilon \sim 10\epsilon _{0}$, one has $L_{D}\sim 0.4\mu m$ and $%
V_{R}=k_{B}T/q=0.0259V$. Therefore for $L=50L_{D}=20\mu m$ one enters the
anomalous regime for $\overline{E}\gtrsim 2kV/cm$. This value of the
electric field is experimentally accessible. In addition, when $\overline{I}%
\gtrsim (L/L_{D})^{2}I_{R}$, that is for $\overline{V}/L=\overline{E}\gtrsim 
\frac{q}{\epsilon }\overline{n}L$, one should enter the regime of shot
noise. For the parameters chosen above we obtain the condition $\overline{E}%
\gtrsim 35kV/cm$ which is still experimentally accessible.\cite{remark4}

In summary, we have proven that a macroscopic diffusive conductor can
display shot noise, and that the transition between thermal and shot-noise
is anomalous when the length of the sample is much longer than the Debye
screening length. The anomaly of the transition consists in a nonlinear
dependence of the low frequency spectral density of current fluctuations
upon the current, which can lead up to a cubic behavior. The origin of this
unexpected behavior is related to the non-trivial coupling among diffusion,
dielectric relaxation and drift in the presence of the long range Coulomb
interaction.


Partial support from the Spanish SEUID and from the EC Improving Human
Research Potential program through contract HPMF-CT-1999-00140, are
gratefully acknowledged.

\begin{figure}[tbp]
\caption{Normalized current spectral density $S_{I}(0)/S_{I}^{0}(0)$ as a
function of the normalized current $\bar{I}/I_{R}$ for different sample
lengths $L/L_{D}=1,10,25,50$, as obtained from present theory, (continuous
lines). For $L<L_{D}$ the curves are indistinguishable from those
corresponding to $L/L_{D}=1$. Also shown for comparison is a standard
crossover between thermal and shot noise for a classical system, as given by
Eq.(\ref{PIintro}) (empty squares) , and the cubic asymptotic expression of
the anomalous crossover as given in Eq.(\ref{PICubic}) (filled circles).}
\label{Fig1}
\end{figure}

\begin{figure}[tbp]
\caption{Characteristic time for number fluctuations normalized to the
dielectric relaxation time $\tau _{N}/\tau_{d}$ as a function of the
normalized current $\bar{I}/I_{R}$ for different sample lengths $%
L/L_{D}=0.1,1$ (dashed lines) and $L/L_{D}=10,25,50$ (continuous lines). }
\label{Fig2}
\end{figure}

\end{multicols}

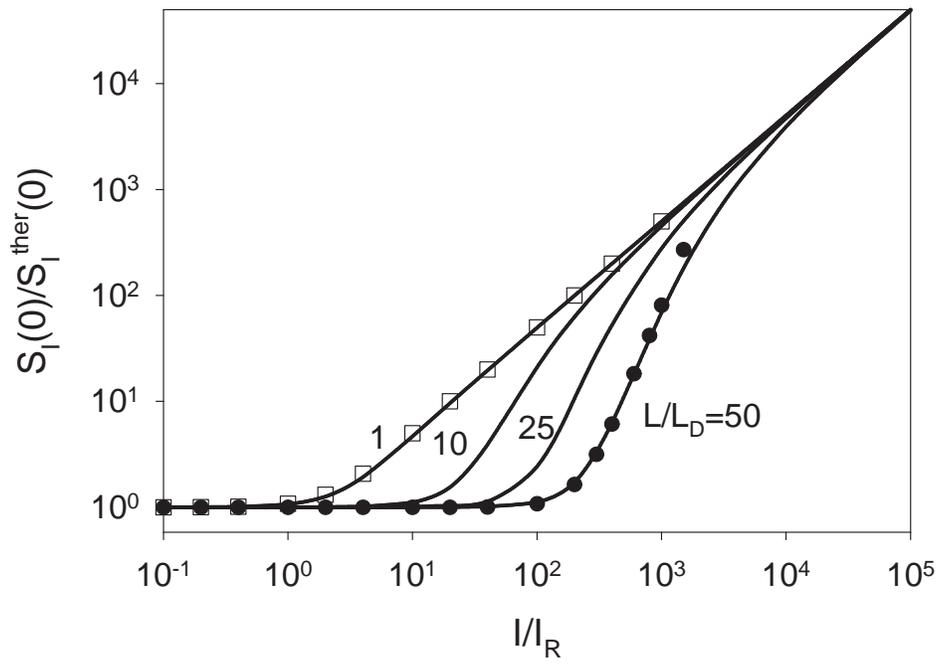
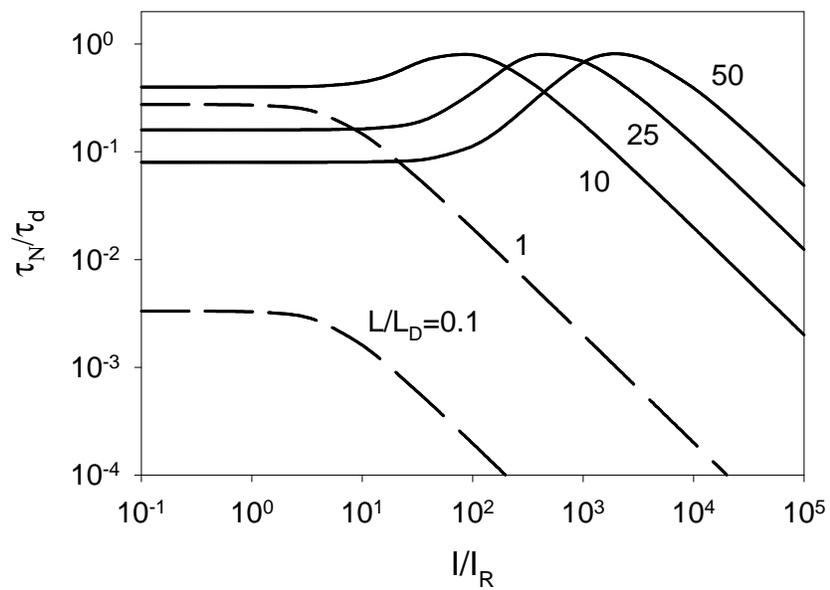
\end{document}